\documentclass{article} % For LaTeX2e
\usepackage{colm2024_conference}

\usepackage{amsmath,graphicx,hyperref}
\usepackage[T1]{fontenc}
\usepackage[utf8]{inputenc}

\usepackage{microtype}
\usepackage{multirow}

\usepackage{xspace,mfirstuc,tabulary}

\usepackage{makecell}
\usepackage{subcaption,siunitx,booktabs}
\usepackage{svg}
\usepackage{array, multirow}
\usepackage{array}
\usepackage{verbatimbox}

\usepackage{listings}
\usepackage{xcolor}
\newcommand{\eat}[1]{}

\newcommand{\name}{\textsc{ROLex}\xspace}
\newcommand{\setting}{\textsc{DKAP}\xspace}

\newcommand{\ie}                {\emph{i.e.},\xspace}
\newcommand{\eg}                {\emph{e.g.},\xspace}

\newcolumntype{M}[1]{>{\raggedright\arraybackslash}m{#1}}

\title{Handling Open-Vocabulary Constructs in Formalizing Specifications: Retrieval-Augmented Parsing with Expert Knowledge}
\author{Mohammad Saqib Hasan$^*$, Sayontan Ghosh$^*$, Dhruv Verma$^*$  
, Geoff Kuenning$^\dagger$, Erez Zadok$^*$,\\ \textbf{Scott A. Smolka$^*$ \& Niranjan Balasubramanian$^*$}\\
$^*$ Department of Computer Science, Stony Brook University\\
$^\dagger$ Department of Computer Science, Harvey Mudd College
\\
\texttt{\{mdshasan, sagghosh, dhverma, ezk, sas, niranjan\}}
\\
\texttt{@cs.stonybrook.edu, geoff@cs.hmc.edu}
}

\colmfinalcopy % Uncomment for camera-ready version, but NOT for submission.
\begin{document}
\maketitle
\begin{abstract}

We study the problem of \emph{Open-Vocabulary Constructs}
(OVCs)---ones not known beforehand---in the context of converting
natural language (NL) specifications into formal languages (e.g.,
temporal logic or code).  Models fare poorly on OVCs due to a lack of
necessary knowledge \emph{a priori}.  In such situations, a domain
expert can provide correct constructs at inference time based on their
preferences or domain knowledge. Our goal is to effectively reuse this
\emph{inference-time},
expert-provided knowledge for future parses without
retraining the model.  We present
\emph{dynamic knowledge-augmented parsing}
(DKAP), where in addition to the input sentence, the model receives
(dynamically growing) expert knowledge as a key-value lexicon that associates NL phrases with correct OVC constructs.
We propose \name, a  \textit{retrieval-augmented
parsing} approach that uses this lexicon.
A retriever and a generator are trained to find and use the
key-value store to produce the correct parse.
A key challenge lies in curating
data for this retrieval-augmented parser.  We utilize
synthetic data generation and the data augmentation techniques on annotated (NL sentence, FL statement) pairs to train the augmented parser. To improve training effectiveness, we propose multiple strategies to teach models to focus on the relevant subset of retrieved knowledge. Finally, we introduce a new evaluation paradigm modeled after the \setting problem and simulate the scenario across three formalization tasks (NL2LTL, NL2Code, and NL2CMD).  Our evaluations show that \setting is a difficult challenge, and \name helps improve the performance of baseline models by using dynamic expert knowledge effectively.\footnote{Code and data are available at \href{https://github.com/StonyBrookNLP/rolex}{\texttt{https://github.com/StonyBrookNLP/rolex}}}

%The evaluations show that IKAP is a difficult challenge and our
%approach is able to effectively use inference-time knowledge to
%address OVC problems.

\end{abstract}

%%%%%%%%%%%%%%%%%%%%%%%%%%%%%%%%%%%%%%%%%%%%%%%%%%%%%%%%%%%%%%%%%%%%%%%%%%%%%%
%% For Emacs:
% Local variables:
% fill-column: 70
% End:
%%%%%%%%%%%%%%%%%%%%%%%%%%%%%%%%%%%%%%%%%%%%%%%%%%%%%%%%%%%%%%%%%%%%%%%%%%%%%%
%% For vim:
% vim:textwidth=70
%%%%%%%%%%%%%%%%%%%%%%%%%%%%%%%%%%%%%%%%%%%%%%%%%%%%%%%%%%%%%%%%%%%%%%%%%%%%%%
% LocalWords:  FSL

%Basic introduction
\section{Introduction}
\label{sec:introduction}

\begin{figure}[!ht]
  \vspace{-1.0em}
  \centering
  \includegraphics[width=0.9\textwidth]{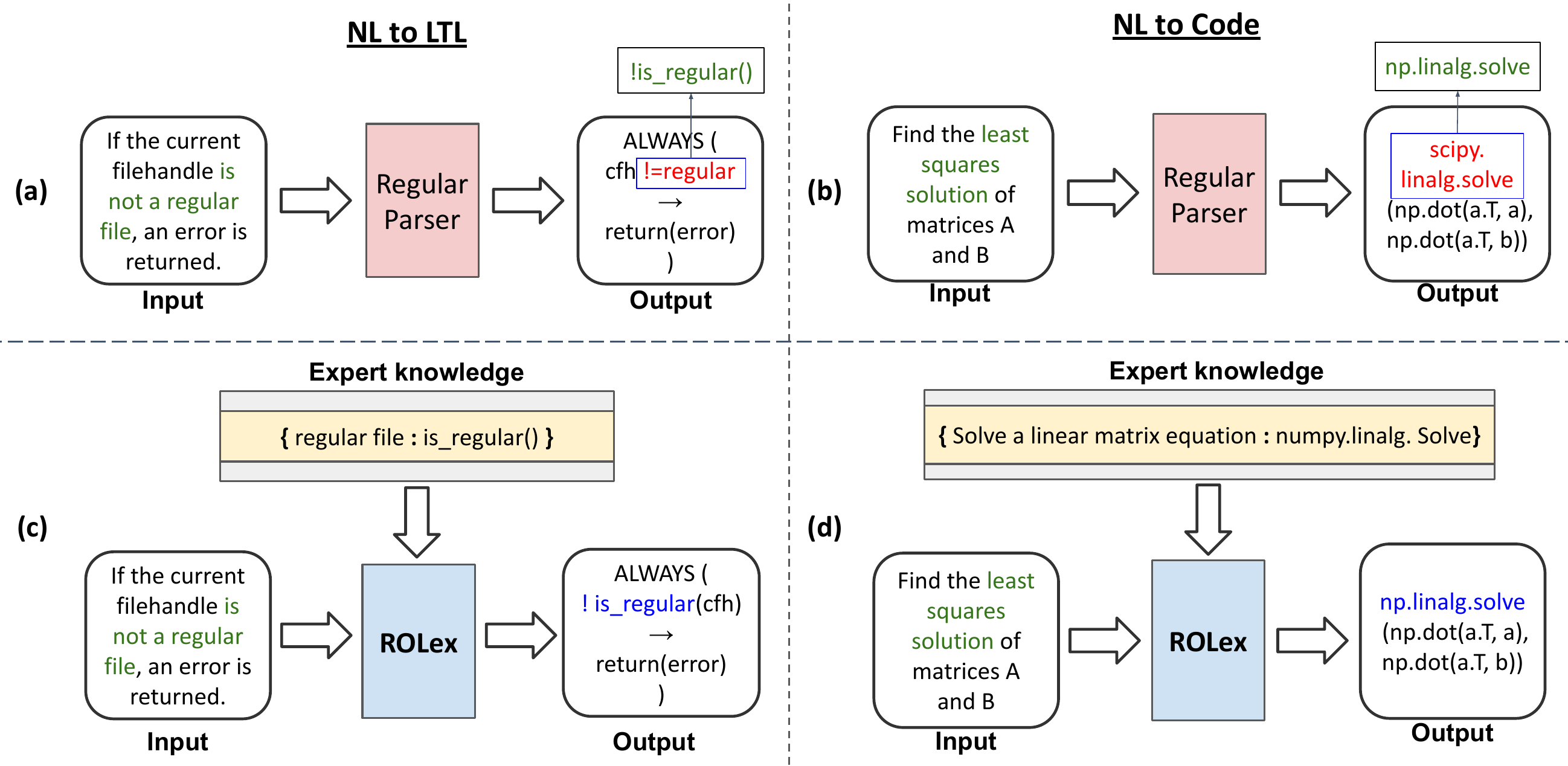}
  \vspace{-1em}
  \caption{ \small{ \textbf{\name in practice}: We show two
      parsing tasks with OVCs: (i) natural language to
      linear temporal logic (NL to LTL) and (ii) natural language to
      code (NL to Code).  As shown in \textbf{(a)} and
      \textbf{(b)}, a regular parser cannot generate the expected
      target constructs because they were not seen in the training data.  The
      phrase \textcolor{teal}{\textsc{is not a regular file}} in
      NL2LTL input should map to
      \textcolor{teal}{\textsc{!is\_regular()}}, but the model
      generates \textcolor{red}{\textsc{!= regular}}.  Similarly for
      NL2Code, the model uses \textcolor{red}{\textsc{scipy.linalg.solve}}
      instead
      of \textcolor{teal}{\textsc{numpy.linalg.solve}} as the library function
      for \textcolor{teal}{\textsc{least squares solution}}.
      \name
      augments models with the ability to use
      expert feedback to handle these OVC cases.  The expert
      knowledge is designed as a key-value lexicon that
      maps NL phrases to their expected parses.
    }
  }
\label{Fig:parking_intro}
\vspace{-2em}
\end{figure}
Formalizing natural language (NL) specifications is a fundamental NLP problem with applications in verification~\citep{nl2tl,deepstl}, code and proof generation~\citep{conala, codeT5, autoformalization}.
% Formalizing natural language (NL) specifications is a fundamental NLP problem with applications in verification~\citep{nl2tl,deepstl}, code generation~\citep{conala, codeT5}, and proof generation~\citep{autoformalization}.
Modern parsers for these tasks are modeled using Seq2Seq~\citep{deepltl, deepmath,deepstl} models, which take the specification as input and output a formal representation as a sequence of n-grams.

%Talk about the challenges in Seq2Seq semantic parsing
A fundamental challenge in NL formalization is the presence of
\emph{Open-Vocabulary Constructs} (OVCs), phrases in the NL
specification for which the formal-language construct is
not known \emph{a priori}. This is due to either poor training data
coverage or a lack
of the domain knowledge needed to anticipate all possible
target constructs. OVC problems are common in formalisms used in verification, such as Linear Temporal Logic (LTL)~\citep{ltl_define} and Signal Temporal Logic (STL)~\citep{deepstl}.  This is due to
phrases that cannot be generalized to a formal construct without prior
knowledge about the specific domain (\ie the software being
verified) or expert preferences.
Even domain experts cannot anticipate all necessary OVC constructs, since they would have to read (parse) each specification sentence, make decisions about how to
represent the constructs and aggregate them into a list, making it a cumbersome
process.  Such constructs are, in effect, unknown to the model and the experts before they are first encountered when parsing the
corresponding specifications.  The NL-to-LTL portion of
Figure~\ref{Fig:parking_intro}
illustrates the use of OVCs with an example from Network File System
specifications~\citep{nfs_rfc}.
OVCs are also prevalent in regular program
synthesis~\citep{conala,
  autoformalization}, as models may use functions from knowledge
that does not meet the downstream runtime environment or user
preference.
The NL-to-Code portion of Figure~\ref{Fig:parking_intro} illustrates
an example where a regular model might use a linear algebra
implementation from the \textsc{scipy} library, whereas the user would prefer a
\textsc{numpy} implementation due to its availability or runtime
requirements.

OVCs present a challenge that is fundamentally different from
previously studied
\underline{o}ut-\underline{o}f-\underline{d}istribution
settings~\citep{docprompting,spider,zerotop}.
OVCs are OOD
problems in that the constructs in the test
specifications are not seen by the model during training.  However, in
the settings we consider, the constructs are also not known or
enumerable a priori: there is no knowledge available to use about
the construct until the first encounter during inference.
This means that we cannot apply techniques relying on
documentation~\citep{docprompting} as a knowledge source for the
construct or existing knowledge bases~\citep{freebase} that provide the
construct vocabulary (\eg predicates in tables).
Other domain-adaptation techniques that use in-domain
examplars~\citep{redcoder,pasupat2021controllable} in retrieval-augmented-generation rely on models' generalization to generate the OVC construct given in-domain
examples.
This generalization, we argue, is a hard challenge for the types of
OVC settings we consider, where deep domain expertise and user
preferences impact construct design and choice.

To address the OVC issue, a model needs feedback from the user about
the correct construct.  In traditional feedback-based frameworks, user
feedback is provided after every parse.  However, in
formalization tasks, we find that OVCs, while not known to the models
during training, repeat multiple times in specification
sentences.  Therefore, OVC feedback from the user can be recorded and
used in a later parse without any fine tuning.
This presents a new knowledge-augmented parsing setup, which we call
\emph{dynamic knowledge-augmented parsing} (\setting).  In regular
knowledge-augmented parsing, a static knowledge database (\eg
Wikipedia, Github) is built before inference~\citep{RAG1,RAG2}.  In
\setting, the knowledge base is dynamic, growing
with each parse.  \setting presents a realistic scenario for solving
the OVC issue because (i)~it is difficult to enumerate all
possible OVCs before they are encountered, (ii)~an expert can generate
knowledge based on errors, and (iii)~OVCs repeat
across parses so knowledge can be reused.

%Describe ParKing

Using \setting, we propose \name, a
\emph{retrieval-augmented generation} (RAG) framework, to help parse
OVCs.  \name collects small amounts of feedback from the
expert based on the errors made by the parser when translating OVCs.
This feedback, designed as a key-value \textbf{lexicon}, represents
knowledge from the expert for the model to use in translating an OVC in future
parses.  The key
is a generalized NL phrase or description
%\ghk{(a) ``/'' is not an English word.  (b) Which is it?  A phrase or a description?}
of an OVC; the value is the correct construct.  A retriever module is
trained to
fetch the relevant set of expert lexicons from the
feedback database during each parse, and a generator module is
conditioned to use this information when parsing.  When an
already-seen OVC is encountered in a future parse, the previously
stored knowledge is used to generate the correct parse.
Figure~\ref{Fig:parking_intro} shows examples of how \name will help
resolve OVCs.  We outline two mechanisms for training \name:
(i)~using synthetic data when there is no training data and
(ii)~using data augmentation techniques when training data exists.  We
also propose several training methods that let the RAG modules learn the
task better.  Furthermore, we show how few-shot LLMs can also benefit
when modeled via \name.

%Describe the efficacy of ParKing
We simulated \name across several benchmarks from various semantic
parsing domains: NL-to-Linear Temporal Logic (NL2LTL), NL-to-Bash
commands (NL2CMD), and NL-to-Python code (NL2Code).  The simulations
were done in the \setting scenario and in the presence of out-of-distribution
constructs to emulate OVCs.  We observed that \name provides consistent
performance gains over baseline seq2seq generation in both fine-tuned
and few-shot settings.  Further analysis shows how each module of
\name is designed to provide optimal performance.

%State the overall contributions of the paper
In summary, this paper makes the following three key contributions:
(1)~introduces a new parsing setting called \emph{dynamic
knowledge-augmented parsing} (\setting) that accounts for growing
user-provided inference-time
knowledge;
%\ghk{It's weird that the italicized phrase doesn't match the acronym.}
%
(2)~presents a retrieval-augmented generation model, \name, and an effective training strategy to address the OVC
problem in semantic parsing in the \setting setting;
(3)~demonstrates the challenges of the \setting problem and the
effectiveness of the proposed solution (\name) through empirical
evaluation and analysis on three different formalization tasks.

\section{Related Work}

% Our work relates to several NLP domains, which we discuss here.

% \paragraph{Semantic Parsing} Research on semantic parsing can be
% categorized to \textit{dataset creation} and \textit{modeling
% parsers}.  Datasets are human curated like
% \cite{dwyer1999patterns}(NL to temporal
% logic),\cite{Gopalan2018SequencetoSequenceLG}(NL to LTL for
% robotics) or synthetic such as ~\cite{text2log,9401852,
% Ontan2022LogicInferenceAN} (text to logic) or
% \cite{deepltl}(Synthetic LTL). Models previously included grammar
% based parsing ~\cite{ccg2lambda,generateVerifByTree} but now uses
% LMs ~\cite{fuggitti2023nl2ltl,nl2tl,
% deepstl,formal_spec_parsing}. Unlike these works, we try to address
% the OVC issue in semantic parsing.

\textbf{Out-of-Distribution (OOD) and Zero-Shot Semantic Parsing:}
Works like Spider~\citep{spider} (Text2SQL) and
Docprompting~\citep{docprompting} introduce out-of-distribution (OOD) semantic parsing
benchmarks. Others develop models to handle OODs.  For example,
SeqZero~\citep{seqzero} decomposes tasks into subtasks, \citep{NQG_T5}
combine model and grammar, \citet{entity_tags} incorporate entity
tags, and \citet{comp_gen_dataset} train LMs to learn syntactic
structure parsing.  Zero/few-shot parsing is also popular owing to low
data requirements.
For instance, Zerotop~\citep{zerotop} decomposes the problem into
subproblems and solves them as zero-shot QA tasks.  In~\citet{autoformalization}, the authors generate theorem code from NL
in zero-shot, while
NL2TL~\citep{nl2tl} generates synthetic data using few-shot prompting
for fine-tuning.  Our work addresses the OVC issue where constructs
are not known \textit{a priori} and thus cannot be resolved before
they are encountered. Hence, unlike OOD settings where there are domain shifts but constructs are known, the need to use constructs never seen
during training makes our OVC  setting much more challenging.
% \nb{Needs a followup sentence here that says why not knowing the constructs a priori makes the previously listed work not directly usable.}

\textbf{Domain Adaptation in Parsing:}
Researchers have proposed domain
adaptation in semantic parsing via retrieval-augmented generation
using exemplars~\citep{pasupat2021controllable} or meta-learning
\citep{chen2020low}.  \citep{shi-etal-2022-xricl} address
cross-lingual Text2SQL with RAG using source domain exemplars for
solving target domain queries, while \citet{nl2tl}, \citet{deepstl},
and \citet{data_recomb} use synthetic data to learn parsing in the
target domain.  Our work also incorporates ideas of domain adaptation
as synthetic data is created to train \name. A key challenge specific
to our work is that all the constructs required to generate the output
are not known apriori. As such, even with synthetic data, generation
models are not exposed to all the possible constructs during training
and need the ability to learn to use new constructs on-the-fly during
the inference time.

% \nb{Saqib and Sayontan, please see the third paragraph in the intro and make sure you include a similar argument here. Simply stating that we also use synthetic training data is a weak connection and doesnt quite say why these existing works are not adequate for our problem.}

\textbf{Retrieval-Augmented Generation (RAG) for Parsing:}
One common RAG for parsing involves retrieving examples from training data~\citep{pasupat2021controllable, gupta-etal-2022-retronlu}.
\citet{zemlyanskiy-etal-2022-generate} introduce an additional step
by doing another retrieval with the input and
generated output and using that as the model input for the final
parse.  \citet{wu-etal-2023-ambiguous} use cyclic training of RAG modules from unsupervised data based on confidence
scores.  Docprompting~\citep{docprompting} uses documentation to
resolve queries about code, while REDCODER~\citep{redcoder} uses
code-comment examples. Our solution is also based on RAG formulation and incorporates expert inputs at inference using a growing knowledge base of lexicons. 
%We address the challenge of effectively training retrievers and generators for this unique setting.
% \nb{The following distinction is not particularly useful. So the setting has a db that keeps growing. Is the RAG problem difficult because of it? Or did we have to do anything different here?}
%regular RAG evaluation where the knowledge database is fixed, \nameworks in a \setting setup where the database grows from scratch during inference.

% \paragraph{Feedback-based Parsing} Feedback-based parsing involves
%models designed to incorporate feedback after output generation,
%either via iterative manner\cite{iterative_feedback} or when model
%prompts for it\cite{model_hil_based_parsing}.  In text-to-code tasks,
%feedback to LLMs helps
%debugging~\cite{ribeiro-lundberg-2022-adaptive}.  \name is designed
%to be a better form of feedback-based parser by reusing expert
%feedback, accumulated during inference, via RAG.

\textbf{OVCs in Semantic Parsing:}
OVCs in parsing previously involved
addressing generalization to new words, for instance, using character
representations~\citep{characterbert, kawakami-learning}.  However, with
modern models, the problem domain has shifted to handling unknown
labels and constructs before they are encountered during
inference. Examples are
\citet{OVC_arg_pred}, which introduces OVC arguments in role prediction
problems, and \citet{OVC_classification}, which introduces
the OVC problem in classification problems with a large number of labels.  
Some other works address OVC predicates 
in semantic parsing, such as \citet{OVC-kb}, where freebase queries are
resolved using knowledge base, and \citet{freebase}, which proposes
predicate matching post generation to handle OVCs. \citet{das-etal-2021} use RAG to address OVC issues by retrieving similar training exemplars and test on out-of-distribution entities in a limited fashion. \citet{mishra-etal-2022} use a dynamic memory of feedback for open-domain QA.  While similar to the above two in using a RAG-based solution to elicit user input, our work differs in terms of the target domain (semantic parsing for specifications versus QA for open-domains), and in its focus on a compact feedback format using key-value lexicons.

\section{Handling OVCs with Dynamic Knowledge-Augmented Parsing}

%This paragraph might go as we already wrote something about it
The \emph{open-vocabulary construct} (OVC) problem refers to the
presence of \emph{a priori} unknown constructs from outside the model's
vocabulary space.  We motivate OVC in two
settings: (i)~formalizing specifications for verifying
complex systems (\eg NL to Linear Temporal
Logic~\citep{ltl_daghstul}) and (ii)~text-to-code (\eg NL to Python code~\citep{conala}).

In formal specifications, the NL constructs are often domain-specific (\ie tied to the
specific software being verified) and thus a generic fixed list is not
particularly useful.  Also, for a given domain, the full set of
needed constructs cannot be enumerated beforehand, since
they can be identified and resolved (by experts) only after reading (parsing) the specification sentences. 
For example, in the Network File Systems domain, there are thousands of network protocol documents (called RFCs), each containing numerous specifications. Experts need to manually parse each such specification with full knowledge of all the possible constructs for building formal models. Experts in our own research group, people with years of experience in formal verification and file systems, struggled to parse these specifications due to the presence of OVCs that they did not know about a priori.
%\sh{For example, experts need to parse thousands of network protocol documents with full knowledge of constructs \emph{apriori} before doing formal verification. In particular, our system domain experts faced difficulty when developing the NL2LTL testbed (details in Section~\ref{sec:task_and_dataset}) due to the OVC issue.}

In text-to-code settings, a user might prefer to use a different implementation (\eg a
preferred Python library) than what the model was trained on,
implement their own method, or use a
specific representation.  Thus, unlike standard semantic
parsing settings, these settings include open-vocabulary constructs,
ones that are unseen or unknown during training, and are only resolved when a specs require construct encountered at test time.

\subsection{Dynamic Knowledge-Augmented Parsing: Problem Definition}

To address this challenge, we introduce the \emph{Dynamic
Knowledge-Augmented Parsing Problem} (\setting), which includes the
following changes to the standard semantic parsing setup:
\textbf{(i)~Dynamic growing knowledge:} When parsing a given test
sentence, the models use a dynamically growing knowledge
base, which records all knowledge provided by the
user for the open vocabulary constructs encountered
in previous sentences.
\textbf{(ii)~Reuse without retraining:} To reduce the user's burden,
we assume that they provide knowledge about a new
construct the first time it is encountered.  The
models must learn to \emph{reuse} this knowledge when the same
construct reappears later.
Appendix Section~\ref{appendix:sec:reuse} empirically motivates the
prevalence of OVCs in a real
specification modeling scenario and their potential for reuse.
Lastly, we want the models to be
trained to use the knowledge as it becomes available,
without having to retrain every time
a new construct is handled.

Consider a stream of NL sentences $X = \langle x_1, \cdots,
x_N\rangle$.  In the standard parsing problem, a model produces an
output $y_t$, conditioning only on the input $x_t$.  In \setting,
when parsing the sentence $x_t$, the model is
additionally given access to expert knowledge $K_{1:t-1}$,
the accumulated knowledge for all open-vocabulary
constructs that appeared in the input sentences $\langle x_1, \cdots,
x_{t-1}\rangle$.  For $x_t$, a model is thus expected to generate the
output $y_t$, conditioning on both $x_t$ and the expert knowledge
$K_{1:t-1}$.

%The model needs to generate $Y = \langle y_1, \cdots, y_N\rangle$
%where $y_t$ is generated by conditioning upon $x_t$ and $K_t$

%Now that we have defined the problem of OVC and partial knowledge,
%let us formally define our \emph{dynamic knowledge augmented parsing}
%setting.

%Assume a stream of NL specifications $X = \langle x_1, \cdots,
%x_N\rangle$ incoming at time $T$.  For a sentence $x_t$ at time
%$T=t$, let the accumulated expert assistance, in the form of expert
%lexicons be

\subsection{Representing Dynamic Expert Knowledge Using Lexicons}

We designed a simple and concise text-to-construct mapping to record
feedback in the form of an \emph{expert lexicon}.  The keys in
the lexicon are generic, idiomatic natural-language phrases; the
values correspond to the correct (\ie desired) formal representations
for the OVCs. Example lexicon entries from two domains, NL2LTL and NL2Code, are
shown in Table \ref{table:model_ipop}. The NL2LTL example, from the
Network File System
(NFS) domain~\citep{nfs_rfc}, shows that the lexicon entry
\textsc{$\langle$A is a regular file
  $\rightarrow$ is\_regular(A)$\rangle$} designating that any phrase
resembling \textsc{A is a regular file} should be parsed to construct
\textsc{is\_regular()} with the argument being the entity
\textsc{A}. Similarly, for NL2Code, the lexicon  $\langle$ Dot product of two
arrays ... $\rightarrow$numpy.dot$\rangle$ indicates that the
dot-product function from the \textsc{numpy} library should be used if
the specification requires it. Formally, when parsing $x_t$, the available knowledge
is given by $K_{1:t-1} = \{(k_i, v_i)\}_{i=1}^{m_{t-1}}$, where
$m_{t-1}$ is the total number of unique OVCs seen until
$x_{t-1}$.  Each key $k_i$ is a generalized natural-language phrase
representing an OVC, and $v_i$ is its corresponding expert-provided
formal representation.

\begin{table}[!t]
    \centering
    \footnotesize
    \begin{tabular}{m{0.08\linewidth}|m{0.27\linewidth}|m{0.26\linewidth}|m{0.29\linewidth}}
\Xhline{3\arrayrulewidth}

 Domain & NL & LEX & FL\\
\Xhline{3\arrayrulewidth}
NL2LTL & If the current filehandle is not a regular file, an error will be returned to the client. & $\langle$current filehandle $\rightarrow$  cfh$\rangle$, $\langle$A is a regular file $\rightarrow$ is\_regular(A)$\rangle$, $\langle$A returned $\rightarrow$ return(A)$\rangle$ &  ALWAYS(( !(is\_regular(cfh)) )$\implies$ ( return(error) ) )\\
\Xhline{2\arrayrulewidth}
NL2Code & 
get the dot product of two one-dimensional numpy arrays
& 
$\langle$ Dot product of two arrays ... $\rightarrow$numpy.dot$\rangle$ 
&
np.dot(a[:, (None)], b[(None), :])\\
\Xhline{3\arrayrulewidth}
\end{tabular}
\caption{\small{Example of natural language specification (NL), relevant expert lexicons needed for parsing (LEX), and a translated formal language statement (FL) from NL to LTL and NL to Code domain.}}
\vspace{-1em}
\label{table:model_ipop}
\end{table}

%%%%%%%%%%%%%%%%%%%%%%%%%%%%%%%%%%%%%%%%%%%%%%%%%%%%%%%%%%%%%%%%%%%%%%%%%%%%%%
%% For Emacs:
% Local variables:
% fill-column: 70
% End:
%%%%%%%%%%%%%%%%%%%%%%%%%%%%%%%%%%%%%%%%%%%%%%%%%%%%%%%%%%%%%%%%%%%%%%%%%%%%%%
%% For vim:
% vim:textwidth=70
%%%%%%%%%%%%%%%%%%%%%%%%%%%%%%%%%%%%%%%%%%%%%%%%%%%%%%%%%%%%%%%%%%%%%%%%%%%%%%
% LocalWords:  FSL OVCs OVC priori NL filehandle LEX cfh

\section{Retrieval-Augmented Parsing with Expert Lexicons}
%\name: Method Description}

\begin{figure*}[!ht]
  \vspace{-1.5em}
  \centering
  \includegraphics[width=0.9\textwidth]{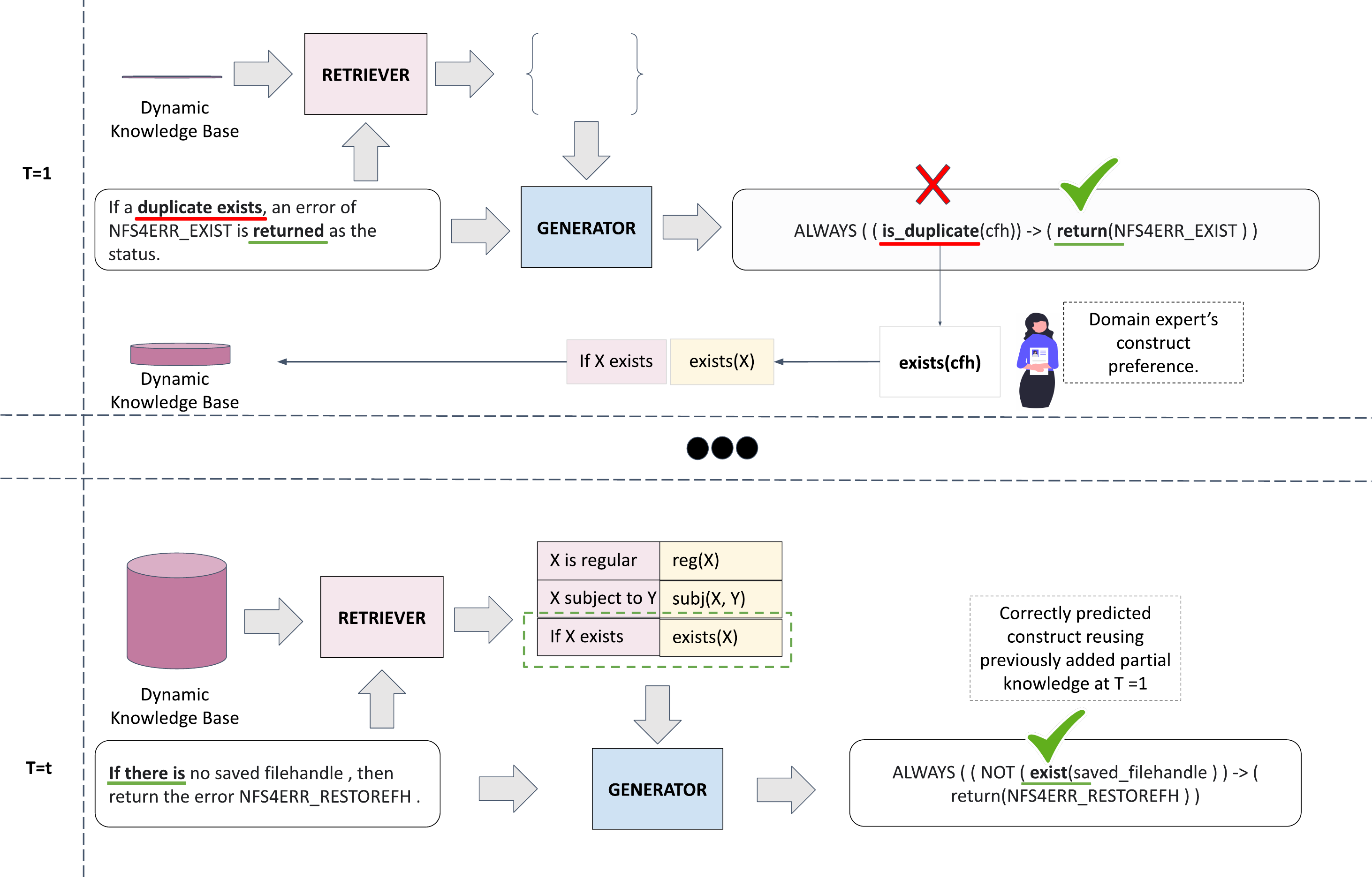}
  \vspace{-0.7em}
  \caption{\small{ \textbf{Dynamic knowledge-augmented parsing using
        \name}: Our RAG framework for inference-time knowledge-augmented parsing through expert lexicons.  We start at $T=1$ with an empty lexicon. The generator thus parses the first NL specification with empty retrieved knowledge to produce a formal statement.
      This generates an error (highlighted in red): ``\textit{duplicate exists}'' is incorrectly parsed as \textsc{is\_duplicate()}, whereas the domain expert prefers ``\textit{exists(X)}''.
      This partial-knowledge entry is added to the dynamic knowledge base and gets reused appropriately at a later time step $T=t$, as shown in the lower block. }
      }
  \label{Fig:parking_example}
  \vspace{-1em}
\end{figure*}

%We described how to address the OVC problem in formalization tasks
%using our designed lexicon in a dynamic knowledge-augmented parsing
%setting.  To effectively model this setup,

To address the \setting setting, we need models that can take an input
NL specification $x_t$ and the current state of the expert lexicon
database $K_{1:t-1}$ at $T=t$ and produce a semantic parse $y_t$ that
includes the appropriate OVC constructs from $K_{1:t-1}$.  A
simple solution is to fit the entire knowledge $K_{1:t-1}$ as
additional context to a generative model to produce the output parse.
This can be problematic for two reasons.  First, the growing knowledge base
may exceed the model's
size limit.  For example, a common input limit for many Seq2Seq models
is 512 tokens, which is, for example, exceeded after aggregating
expert lexicons from only 25 NFS RFC
specifications.  Second, adding the entire $K_{1:t-1}$ may increase
the \emph{noise} (\ie irrelevant entries) in the input
context for the generator.  A retriever helps mitigate this problem by
selecting the relevant parts of $K_{1:t-1}$ as input.

Therefore, we propose \name, \underline{R}etrieval-augmented generation parsing with expert-provided \underline{O}VC \underline{lex}icons, illustrated in Figure~\ref{Fig:parking_example} using examples from the NFS RFC. We can fine-tune LMs such as T5~\citep{t5} or Code-T5~\citep{codeT5}, or few-shot prompt LLMs like ChatGPT~\citep{chatgpt} and GPT-4~\citep{gpt4}) for the generator module, and use dense retriever models~\citep{simcse,dpr} for the retriever.  The main challenge lies in obtaining training data.  To address this, we use synthetic data generation and data-augmentation techniques. To develop the retriever and generator modules, we experimented with several methods, explained next.

%----------------------------------------------------------------------
\subsection{Dataset Creation and Augmentation}
\label{sec:dataset_creation}

To train the generator, we need training instances of the form
$(x,K,y)$; to train the retriever, we need instances of the form
$(x, K, K_y)$, where $K_y \subseteq K$ is the subset of relevant
key/value pairs for the OVCs that appear in the output $y$ for
instance $x$.  Here we describe our approach for generating and
augmenting such training data automatically.

\textbf{Synthetic Data Generation:}
In some semantic parsing use cases, such as formalizing RFC
specifications, manually curating $(x,y)$ pairs and lexicons for OVCs can be difficult.
% and expensive.
However, due to
the restricted nature of the parses, we can
% easily
curate synthetic
data via grammar-based generation~\citep{deepstl}.  We do this by
developing a synthetic grammar $\mathcal{G}$ adapted to the target
domain.  For example, we have shown that such a grammar can be used to model LTL statements generated from the NFS domain.
Details of the grammar are provided in the Appendix Section~\ref{appendix:sec:grammar}.

\textbf{Data Augmentation Setting:} If training data with $(x,y)$ pairs exist, we can augment them with the gold lexicon $K$ for each pair. We do this by first defining the OVCs and their formal representations, and then using a suitable pre-defined knowledge source to obtain an idiomatic natural-language description of the OVCs.  For example, the first few words from Python documentation serve as OVC descriptions in the NL2Code domain.  In this way, we construct a comprehensive lexicon knowledge base and then use it to enumerate each $(x,y)$ pair with the relevant $K$.

%----------------------------------------------------------------------
\subsection{Developing the RAG Modules}
\label{sec:training}

\textbf{Retrieval Module:}
We use sentence transformers~\citep{sbert} as our dense retrievers in two ways:
off-the-shelf and fine-tuning.
For \emph{off-the-shelf} usage, we plug in a pre-trained
retriever.  However, this approach does not work well in
formalization settings because most retrievers are fine-tuned on general NL tasks, and each parsing domain has its unique
language and set of sentences to model.
Hence, we opt for \emph{fine-tuning the retriever} using our
curated $(x,K,y)$ data pairs.  Given $K = \{(k_i, v_i)\}_{i=1}^{m}$,
where $k_i$ is the natural-language grounding phrase and $v_i$ is the
OVC's formal representation, we create ${(x,k_i)}_{i=1}^{m}$ tuples
for each $x$.  This tuple store is then used to fine-tune a retriever $\mathcal{R}$
using contrastive learning~\citep{multiple_negatives_ranking_loss} with
in-batch negatives~\citep{dpr,chenetal} (see
the Appendix Section \ref{appendix:sec:retriever_loss}, for the
training objective). The retriever $\mathcal{R}$ is fine-tuned to include all $K$ gold lexicon entries in its top $n$ ranked lexicon entries.

\textbf{Generator Module:} The Generator $\mathcal{G}$ is a language model that should generate the semantic parse $y$
based on $(x,K^*)$, where $K^*$ is the list of top $n$
ranked lexicon entries retrieved by $\mathcal{R}$ from the full lexicon.
Ideally, the generator should be able to identify $K_y$, the relevant lexicon entries in $K^*$ and then use them appropriately to produce the correct semantic parse $y$. However, in practice, there can be three types of issues that arise when training generators with noisy retrieved lexicon entries. The generator might:
(i)~entirely ignore $K^*$ during generation, (ii)~use non-relevant entries in $K^*$, resulting in incorrect parses, and (iii)~fail to identify or use \emph{all} of the relevant lexicon entries.
To mitigate these issues, we propose four generator training schemes:
\textbf{(1)} \textsc{Basic}: $\mathcal{G}$ is trained by
generating $y$ conditioned on $x \bigoplus K^*$.
\textbf{(2)} \textsc{Extra Supervision}: From each $(x,K^*,y)$
training pair, we derive two pairs: $(x,K^*,y)$ and
$(x,K_y,y)$.  This causes $\mathcal{G}$ to observe $y$ during training
twice---once conditioned on $x \bigoplus K_y$ and another time on $x
\bigoplus K^*$.  The idea is to provide an extra supervision signal to
$\mathcal{G}$ on what are correct lexicons.
\textbf{(3)} \textsc{Multi-task Learning}: $\mathcal{G}$ is
trained to generate $k_y \bigoplus y$ conditioned on $x \bigoplus K^*$.
The motivation lies in workings of language models: they generate the
next token based on the input and the prior generation.
\textbf{(4)} \textsc{Transfer Learning}: $\mathcal{G}$ is
trained first to generate $k_y$ conditioned on $x \bigoplus K^*$.
Then model is further fine-tuned as in scenario (1). Intuition is that relevant lexicon extraction helps $\mathcal{G}$ learn to parse better.

\section{Evaluation Setup}

% Then, we synthetically generate and augment training and indexed
% test sets $\mathcal{D}$ of tuples $(x_t,K_t,y_t)$.

% \sg{Saqib, it is hard to parse this sentence}
Our goal is to evaluate \name for addressing OVC
issues in the Dynamic Knowledge-Augmented Parsing (\setting)
setting. We
introduce a new evaluation protocol that differs from the standard
evaluation for RAG: (i)~the knowledge
base grows as we parse more specifications and (ii)~an expert adds new
knowledge to correct OVC errors after each parse.

%Look at it again later on
\textbf{\setting Evaluation Protocol:} Our evaluation protocol mimics
dynamic (growing) knowledge of OVCs at test time by simulating the
addition of knowledge about a new OVC after each parse. In this way,
subsequent parses can use this accumulated knowledge. Formally, given
task $\mathcal{T}$ and possible OVCs $O$, we create train and test
sets, $\mathcal{D}_\mathit{train}$ and $\mathcal{D}_\mathit{test}$ as
per methods in Section~\ref{sec:dataset_creation}. Each $\mathcal{D}$
consists of tuples $(x_t,K_t,y_t)$ where $x_t$ is the NL, $y_t$ is the
FL and $K_t$ are the gold lexicons. \name is trained on
$\mathcal{D}_\mathit{train}$ and then evaluated with an initially
empty dynamic knowledge base, $\mathcal{KB}$. At time $t$, the
retriever retrieves relevant lexicons $K^*$ from $\mathcal{KB}$, and
$x\bigoplus K^*$ is given as input to generate $y'$. Comparing $y'$
and $y$, we extract and add a set of new lexicons $K_\mathit{new} \in
K_t$, s.t. $K_\mathit{new} \cap \mathcal{KB} = \emptyset$ to
$\mathcal{KB}$ so that they can be retrieved for parsing at time
$t+1$.

\subsection{Task and Dataset Details}
\label{sec:task_and_dataset}

We evaluated \name across three formalization tasks. Our
evaluation sets were created with constructs in the target output that
were not present in the training sets, thus simulating out-of-vocabulary
constructs (OVCs). The first task, \textbf{NL2LTL}, was to parse NFS
specs into Linear Temporal Logic (LTL); the OVCs are function
predicates (verbs) and a subset of variables (nouns). Due to lack of
training data, we opted for grammar-based synthetic data generation
(outlined in Section \ref{sec:dataset_creation}). Details about
the grammar are given in Appendix Section \ref{appendix:sec:grammar}. Our test
set consisted of $100$ NFS statements taken from
SpecNFS~\citep{specnfs} and annotated to LTL by experts. Our second
task is \textbf{NL2Code}, which involves formalizing specs to Python
code, and the OVCS are function names. We use the CoNaLa dataset from
Docprompting as the test set that contains unseen functions to simulate
OVCs. We augment the data with lexicons using the database of Python
documents provided with the dataset. Our final task, \textbf{NL2CMD},
involves converting statements to bash commands where the OVCs are bash
functions. Here, we again use TLDR dataset from Docprompting as the test
set that contains OOD commands. To create the lexicon, we utilize
tldr\footnote{https://tldr.sh/} pages to form lexicon keys. Further
details and examples from each dataset, \textit{along with information on how the lexicons were designed}, are listed in Appendix Section ~\ref{sec:appendix:dataset_details}.

\subsection{\name Benchmarking Setup}
\label{sec:benchmarking_setup}
We used the different (small, base, large) variants of the BGE
sentence transformer from FlagEmbedding~\citep{flagembedding} as our
retriever. For the generator, we used T5 and
Code-T5 for the fine-tuned setting and
ChatGPT and GPT-4 for few-shot. For
evaluation, we calculate BLEU~\citep{bleu} to measure quality of whole
parse and precision, recall, and F1-score of the OVCs to measure
performance on OVCs. We also calculate the gains \name achieves using
reusable feedback compared to traditional feedback by measuring
reduction in error. Further details are given in Appendix Section
\ref{appendix:sec:setup_and_eval}.

\section{Results \& Analysis}
Our experimental evaluation setup is aimed at answering the following questions: 
(i)~How effective is \name in leveraging expert knowledge to address
OVCs? (Tables~\ref{table:main}  and~\ref{table:few-shot_and_retriever}(a))
(ii)~What is the effect of retriever performance on the overall
downstream task performance? (Table~\ref{table:few-shot_and_retriever}(b)). 
(iii) What is the variation in the overall OVC task performance for
different generator training approaches?
(Table~\ref{tbl:generator_and_lexicon}(a)). 
(iv) What is the effect of the format of expert knowledge on the OVC
task? (Table~\ref{tbl:generator_and_lexicon}(b)). Note all our evaluations are conducted on testbeds containing OVCs unseen during training to reflect the actual nature of the setup.
% How much does the trained retriever contribute to \name's performance (b))? (iii)~How effective is our generator training strategy (Table~\ref{tbl:generator_and_lexicon}(a))? (iv)~How effective is the lexicon based knowledge (Table~\ref{tbl:generator_and_lexicon}(b))?
% In all evaluations, pre-trained LLMs are fine-tuned or prompted in few-shot settings. Baselines use the $(x, y)$ pairs in each dataset. Corresponding \name versions use the $(x, K^*, y)$ for training/few-shot and tested using $(x, K^*, y)$ following the IKAP evaluation protocol. 

% \subsection{Results on Benchmarks}
\subsection{Results with fine-tuned LLMs}
Table~\ref{table:main} reports \name's performance when different
pre-trained LMs are fine-tuned as generators for the three different OVC
tasks. For each model, we compared a baseline generator that parses FL
from NL without additional context against the corresponding \name
generator. All generators were trained using the \textsc{Transfer
  Learning} method, where we first train the model to extract lexicon
followed by training it on the downstream task (see
section~\ref{sec:training}).
For the retriever, we used the large variant of the BGE model, which
was fine-tuned for the lexicon retrieval. Without dynamic knowledge,
models fare poorly on the OVCs across all three tasks. The results
also show that \name yields substantial improvements on the OVCs as
well as on the overall generation, when compared to the baseline that
does not use any additional knowledge. In NL2LTL, for example, the
gains range from $20\%$ to $38\%$ in OVC F1. These improvements show that \name uses the dynamic knowledge during inference to address the OVC issue.

\begin{table*}
\centering
\small
\resizebox{\linewidth}{!}{%
\begin{tabular}{m{0.08\linewidth}|m{0.04\linewidth}|m{0.04\linewidth}|m{0.04\linewidth}|m{0.04\linewidth}|m{0.04\linewidth}|m{0.04\linewidth}|m{0.04\linewidth}|m{0.04\linewidth}|m{0.04\linewidth}|m{0.04\linewidth}|m{0.04\linewidth}|m{0.04\linewidth}}

\Xhline{3\arrayrulewidth}
\multirow{3}{*}{\makecell[l]{Model}}& \multicolumn{4}{c|}{NL2LTL} & \multicolumn{4}{c|}{NL2Code}& \multicolumn{4}{c}{NL2CMD}\\\cline{2-13}

& \multicolumn{1}{c|}{\makecell[c]{All}} & \multicolumn{3}{c|}{\makecell[c]{OVCs}} &  \multicolumn{1}{c|}{\makecell[c]{All}} & \multicolumn{3}{c|}{\makecell[c]{OVCs}}&  \multicolumn{1}{c|}{\makecell[c]{All}} & \multicolumn{3}{c}{\makecell[c]{OVCs}}\\

\cline{2-13}
& \multicolumn{1}{c|}{bleu} & Prec. & Rec. & F1 & \multicolumn{1}{c|}{bleu} & Prec. & Rec. & F1 & \multicolumn{1}{c|}{bleu} & Prec. & Rec. & F1\\
%\Xhline{3\arrayrulewidth}
\hline
T5-B & 27.94 & 39.35 & 43.42 & 41.28  &
18.51 & 4.89 & 4.84 & 4.86 &
13.17 & 4.96 & 7.22 & 5.88 \\
+\name &
53.28 & 68.67 & 55.75 & 61.54 &
23.02 & 18.85 & 20.50 & 19.64 &
15.44 & 25.47 & 37.39 & 30.30 \\
\hline
CT5-S & 30.59 & 34.83 & 24.75 & 28.94 &
16.90 & 5.68 & 5.13 & 5.39 &
18.10 & 3.47 & 5.39 & 4.22 \\

+\name &
42.95 & 62.42 & 43.33 & 51.15  &
25.81 & 15.93 & 16.17 & 16.05 &
22.27 & 29.10 & 37.18 & 32.65 \\
\hline
CT5-B & 25.15 & 21.33 & 14.08 & 16.97 &
29.30 & 14.98 & 14.30 & 14.63 &
26.04 & 2.46 & 4.31 & 3.13 \\
+\name &
54.01 & 63.83 & 48.00 & 54.80 &
31.81 & 17.35 & 18.85 & 18.07 &
27.11 & 20.83 & 32.00 & 25.24 \\

\Xhline{3\arrayrulewidth}

\end{tabular}
}
\caption{\small{
Performance gains in fine-tuned models across three different
benchmark tasks due to \name. \texttt{Bleu} in \texttt{All} was
evaluated for the complete model output expression, whereas
\texttt{OVC's} \texttt{F1}, \texttt{Precision} \& \texttt{Recall} were
evaluated only for the subset of OVC constructs in the target
expression.
% 
% Table of results across all benchmark datasets for each of the different fine-tuned models
}
}
\label{table:main}
\vspace{-1em}
\end{table*}

\subsection{Results with Few-Shot LLMs}

Table~\ref{table:few-shot_and_retriever}(a) shows the results of few-shot LLMs
as generators in three settings: (i) no external knowledge is
provided, (ii) bm25 retriever~\citep{bm25} is used to retrieve
external knowledge, and (iii) fine-tuned bge-large retriever (\name)
is used to retrieve the external knowledge. We provide three few-shot
examples.
We observe consistent gains in both OVC F1 and the overall target generation BLEU for \name against baseline and when using an bm25 retriever. 
GPT-4 + \name gave the best performance on the dataset, even against fine-tuned
versions. 
However, we observe that the best fine-tuned \name (T5-base) from Table~\ref{table:main} performed better than both few-shot LLM baselines. 
While adding ROLex to GPT-4 performs the best, we find that T5-base
with ROLex has similar performance to that of ChatGPT with
ROLex. These results suggest the potential of smaller models
fine-tuned with \name when compared to significantly larger models.

\begin{table*}%
\hspace{-2cm}
\begin{subtable}{0.75\linewidth}
\small
\centering
  \begin{tabular}{m{0.11\linewidth}|m{0.06\linewidth}|m{0.06\linewidth}|m{0.06\linewidth}|m{0.06\linewidth}}
\Xhline{3\arrayrulewidth}
\multirow{2}{*}{\makecell[l]{Model}} & \multicolumn{1}{c|}{\makecell[c]{All}} & \multicolumn{3}{c}{\makecell[c]{OVCs}}\\
\cline{2-5}
 & BLEU & Prec. & Rec. & F1\\
% \Xhline{3\arrayrulewidth}
\hline
ChatGPT & 42.23 & 56.70 & 38.14 & 45.61\\
+bm25 & 51.27 & 67.18 & 50.00 & 57.33\\
+\name & 53.03 & 71.43 & 53.26 & 61.02\\
\hline
GPT-4 &  46.98 & 61.86 & 39.60 & 48.29\\
+bm25 & 62.10 & 79.64 & 70.02 & 74.52\\
+\name &  66.38 & 78.87 & 75.43 & 77.11\\
  \Xhline{3\arrayrulewidth}
\end{tabular}
\caption{Few-shot model performance}
\end{subtable}\hspace{-3.5cm}
\begin{subtable}{0.75\linewidth}
\small
\centering
\begin{tabular}{m{0.18\linewidth}|m{0.06\linewidth}|m{0.06\linewidth}|m{0.06\linewidth}|m{0.06\linewidth}}
\Xhline{3\arrayrulewidth}

\multirow{2}{*}{\makecell[l]{Retriever}} & Retr.& \multicolumn{3}{c}{\makecell[c]{OVCs}}\\
\cline{2-5}
 & R@10 & Prec. & Rec. & F1 \\
%\Xhline{3\arrayrulewidth}
\hline
Off-the-shelf & & & & \\
-bm25 & 2.59 & 10.88 & 10.44 & 10.66 \\
-small & 17.91& 13.64 & 13.59 & 13.62\\
-large & 18.81&  14.20 & 14.62 & 14.40 \\
\hline
Fine-tuned & & & & \\
-small & 19.37& 15.30 & 15.33 & 15.31 \\
-large & 20.38& 17.35 & 18.85 & 18.07\\
\Xhline{3\arrayrulewidth}
\end{tabular}
\caption{Effect of retriever performance on downstream task}
\end{subtable} 
\caption{\small{\textbf{(a)} Few-shot model performance on the NL2LTL
  task. Augmenting with \name showed consistent performance
  improvements across all of the metrics. \textbf{(b)}~Effect of
  retriever performance (top@k recall) on the NL2Code task with the best-performing model Code-T5-base.}}
\label{table:few-shot_and_retriever}
%\vspace{-1em}
\end{table*}

\begin{table*}%
\hspace{-2cm}
\begin{subtable}{0.77\linewidth}
\small
\centering
\begin{tabular}{m{0.21\linewidth}|m{0.06\linewidth}|m{0.06\linewidth}|m{0.06\linewidth}|m{0.06\linewidth}}
\Xhline{3\arrayrulewidth}
 \multirow{2}{*}{\makecell[l]{Training\\Strategies}} & All & \multicolumn{3}{c}{\makecell[c]{OVCs}}\\
\cline{2-5}
& BLEU & Prec. & Rec. & F1 \\
\hline
%Nl2Code (CT5-B) &  & &  &  \\
\textsc{Basic (1)} & 29.94 & 16.56 & 17.30 & 16.92  \\
\textsc{Extra Sup. (2)} & 27.69 & 15.77 & 16.80 & 16.27  \\
\textsc{Multitask (3)} & 26.53 & 18.72 & 21.21 & 19.89\\
\textsc{Transfer (4)} & 31.81 & 17.35 & 18.85 & 18.07\\
\Xhline{3\arrayrulewidth}
\end{tabular}
\caption{Effect of training strategies}
\end{subtable}\hspace{-3cm}
\begin{subtable}{0.75\linewidth}
\small
\centering
\begin{tabular}{m{0.11\linewidth}|m{0.07\linewidth}|m{0.07\linewidth}|m{0.07\linewidth}|m{0.07\linewidth}}
\Xhline{3\arrayrulewidth}

 \multirow{2}{*}{\makecell[l]{Feedback}} & All & \multicolumn{3}{c}{\makecell[c]{OVCs}}\\
\cline{2-5}
& BLEU & Prec. & Rec. & F1 \\
\hline
\textsc{Exmp} & 33.33 & 14.98 & 15.96 & 15.46 \\
\hline
\name & & & & \\
-\textsc{Docs}  & 29.54 & 15.69 & 15.30 & 15.49\\
-\textsc{Lex} & 31.81 & 17.35 & 18.85 & 18.07 \\
\hline
\Xhline{3\arrayrulewidth}
\end{tabular}
\caption{Evaluating different feedback mechanisms}
\end{subtable} 
\caption{\small{\textbf{(a)} Effect of generator training strategy on the downstream task of NL2Code, with BGE-large as the retriever. \textbf{(b)}~Evaluating different feedback mechanisms for RAG.}}
  \label{tbl:generator_and_lexicon}%
\vspace{-1em}
\end{table*}

\subsection{Analysis}

We analyzed the different components of \name to justify our design
choices. All analyses were conducted on the NL2Code benchmark using the
Code-T5 base as the \name generator. Finally, we studied the potential
gains of \name for each benchmark.

\textbf{Retriever Analysis}:
Table~\ref{table:few-shot_and_retriever}(b) reports the top-10 recall (R@10) performance of different retrievers, and the precision, recall and F1 for the OVCs in the downstream task. 
% Table~\ref{table:few-shot_and_retriever}(b) reports the performance of different retrievers for their top-10 recall (R@10), and on the downstream task through precision, recall and F1 for the OVCs. 
We tried bm25, BGE-small and large retrievers, both with and without
% We tried bm25, and BGE-small and large retrievers, both with and without
fine-tuning. We observe that dense retrievers (BGE) outperform BM25 by
at least $15\%$ on retriever recall. Fine-tuning and increasing model
size also lead to improved performance. We can also see that improved
retriever performance translates to better downstream task scores. As
the best retriever performance is quite low, improving retriever
performance is a direction worth pursing to do well on the OVC parsing
task. Further details are given in Appendix Section
\ref{appendix:sec:retriever_ablations}.
% \nb{The R@10 numbers seem low to me. Why are they so low? This means that only in 20\% of the cases we are finding the relevant knowledge within the top 10 entries. I dont know if it will hurt us to call attention to this. How do you compute Recall for instances where there relevant OVC is not in the entire lexicon at all. Maybe you cant do much about this at this point. Ignore if there isnt' anything useful to say.}
%This observation justifies the reasoning behind fine-tuning the retriever.

\textbf{Generator Training Strategies}:
We saw that retrieving the relevant lexicon is challenging, so we want generators that can generate the target based on the noisy retrieved knowledge. For this, we evaluated four different generator training strategies: 
% 
% The retriever helps to reduce noise in the selected knowledge. However, the generator must also learn to operate over the potentially noisy retrieved knowledge.
% We compare different training strategies aimed at getting the generator to focus on the relevant knowledge in 
\textbf{\textsc{BASIC}} Directly using the retrieved knowledge to train the generator.
\textbf{\textsc{EXTRA SUP.}} Using extra supervision by mixing in instances with only the gold knowledge.
\textbf{\textsc{MULTITASK}} Getting the model to focus on the relevant knowledge via \textit{multi-task learning}.
\textbf{\textsc{TRANSFER}} Getting the model to focus on the relevant
knowledge via a \textit{transfer learning} approach.
Model performance on these four approaches is reported in Table~\ref{tbl:generator_and_lexicon}(a). 
From the table we observe that (\textbf{\textsc{Approach~1}, 2}) are
not up to the mark, whereas multi-task and task-transfer learning
show substantial improvement over the first two. Based on the
experimental results we choose the transfer-learning approach as our go-to
generator training approach as it provides the best balance between
the OVC F1-score and overall parse quality.

% fares poorly. Using extra supervision by mixing in instances with only the gold knowledge (\textsc{Case~2}) also fares poorly. Getting the model to focus on the relevant knowledge via Multi-task training (\textsc{Case~3}) or in a Transfer Learning setting (\textsc{Case~4}) both yield improvements. We chose the transfer learning setup since it gave the best balance between the OVC F1 and overall parse quality in terms of BLEU.

%As Table~\ref{tbl:generator_and_lexicon}(a) shows that \textsc{Case~3}, in which the generator is first trained to recognize the correct lexicon entries for the input sentence, and then switched to the target parsing task, works the best. 

%We show the OVC metrics when the generator was trained using the different strategies enumerated in Section~\ref{sec:training}, Table~\ref{tbl:generator_and_lexicon}(a). Observing the values, we see that \textsc{Case~3} gave the best OVC performance but poor generalization on all constructs. This indicates, in this scheme, that the generator learned to extract the OVC from the lexicon but overall generation was poor. \textsc{Case~4}, which was used in all our experiments, provided the best tradeoff between OVC and overall generalization.

\textbf{How useful is the lexicon based encoding of knowledge?}:
In Table~\ref{tbl:generator_and_lexicon}, we analyze different forms
of retrieval context to help models overcome OVC issues. We trained
generators to use exemplars~\citep{redcoder} (\textsc{Exemplars}) and
documentation~\citep{docprompting}, the lookup key in our lexicon, (\textsc{Docs});
(see Appendix Section \ref{appendix:sec:feedback} for details).  The
results show that while exemplars have better overall BLEU, \name with
lexicons (\textsc{Lex}) gets the best performance on
OVCs.
%This is because, while (spec, code) examples give models signal to write better code, they do not give models any hint on how to resolve OVCs. 
This is because the (spec, FL) exemplars only provide generic signals
about good parses and not specific information about how to represent
the OVCs in the test sentence.
Also, lexicons are better than just documentation as context, in part due to the succinct description of the relevant
information needed to connect the NL to the corresponding FL
construct.

%This could be due to the lexicon explicitly providing the construct name for the model to use.

\textbf{Estimated Reduction in Expert Effort}
In a standard feedback system the expert has to read every output and correct each erroneous OVC. \name aims to reduce this
burden on the expert by making the feedback reusable. We calculated
the maximum possible percentage gains in saving human effort (treating
reading and correcting entries as unit-cost actions) for each of the
fine-tuned benchmarks and for the
few-shot scenario. We observe $24.3$\% savings for NL2TL, $5.6\%$
for NL2Code, $16.3\%$ for NL2CMD, and $32.9\%$ for few-shot
learning in NL2LTL. These results indicate that dynamic user feedback is a
fruitful direction worth pursuing in settings where the OVC problem is
prevalent. Please see Appendix Section
\ref{appendix:sec:gains} for more details on these results.

\textbf{Error Analysis of OVCs}: 
For the NL2LTL task, we analyzed the OVC errors for $25$ T--base
generations with the lowest BLEU scores. We observed that $23.3\%$ of
the errors were due to relevant expert knowledge in the retrieved
lexicons not being used for generation.
About $23.3\%$ of the errors can be attributed to incorrect lexicon from the set being used.
Almost $46.7\%$ incorrectly generated OVCs were due to  generator hallucination.
For NL2Code we find that $56\%$ are cases where relevant retrieved
lexicons are not used, $28\%$ are cases where relevant lexicons are
not retrieved and $16$\% where an incorrect lexicon is used. This
analysis shows that \name's ability to use lexicons and not
hallucinate are key issues to address for better performance. Further
details are given in Appendix Section \ref{appendix:sec:error_analysis}.

\section{Conclusion \& Future Work}

Formalizing specifications requires expert feedback because of OVCs not resolvable \emph{a priori} without domain
knowledge.  In this work, we propose a new dynamic knowledge-augmented parsing setup and a
RAG model to tackle the OVC problem.  We
show how our approach addresses the OVC issue by accumulating small, reusable expert feedback as a lexicon.
Furthermore, we analyzed different training schemes to develop our model.  Evaluations on multiple
benchmarks highlighted the challenge of our new parsing setting and
demonstrated our approach's ability to address it. %\sh{Addressing open-vocabulary constructs is critical for formalizing the specifications of complex systems. We hope our new setup attracts researchers, with works on improved modeling and better user-centric evaluations of feedback-based parsers.} 
However, there is further room for improvement, especially in developing models that can better utilize expert knowledge, and are generalizable to more domains both in terms of generation and retrieval. We also hope our new problem setup spurs future research not only on improved modeling but also on user-centric evaluations of feedback-based parsers.

\subsubsection*{Acknowledgements}
We thank the anonymous reviewers of COLM for their feedback which helped improve the paper. This work was possible thanks to the NSF award CCF-1918225.
%\sh{We thank the COLM reviewers for their valuable feedback. This work was possible thanks to the NSF award CCF-1918225.}  

%s%This work is based upon research that has been supported by the National Science Foundation under award FMitF $\#1918225$. The authors would like to thank the anonymous reviewers and the area chair for their feedback that helped improve this work. 

% Wethank the ACM HotStorage anonymous reviewers and our shepherd Ram Alagappan for their helpful feedback. This work was made possible in part thanks to Dell-EMC, NetApp, and IBM support; and NSF awards CCF-1918225, CNS-1900706, CNS-1900589, CNS-1729939, CNS-1730726, DCL-2040599, and CPS-1446832.

% % \input{sections/results_new}
% \input{sections/conclusion}
% \input{sections/limitations}
\bibliography{colm2024_conference}
\bibliographystyle{colm2024_conference}
\newpage
\section*{Appendix}
\appendix

\section{Grammar for synthetic generation}
\label{appendix:sec:grammar}

Figure \ref{table:grammar} shows the custom grammar designed to model
and generate the synthetic NL2LTL dataset. We can generate example
statements in the language defined by this synthetic CFG by
stochastically chaining the production rules beginning with the start
symbol. Since the natural-language phrases are also attached to the
grammar, the process simultaneously yields the corresponding NL
statement $x$ for the generated formal statement $y$. We can create a
large number of such $(x, y)$ pairs by sampling from this grammar. To
simulate the expert-provided knowledge, we augment each pair with a
knowledge set $K$ that contains a random collection of key/value pairs
as distractors, and also the relevant key/value pairs that correspond
to specific OVCs appearing in the output $y$. We can thus control/vary
how the distractors are created.

\begin{figure}[!ht]
\begin{tabular}{l l l l}
$1.$ & <start> & $\rightarrow$ & G(<phrase>)\\
$2.$ & & $\rightarrow$ & G(<phrase> $\implies$ <phrase>)\\
$3.$   &     & $\rightarrow$ & G(<phrase> $\implies$ <phrase> $\vee$\\
    &    & & $\neg$ <phrase> $\implies$ <phrase>)\\
$4.$ & <phrase> & $\rightarrow$ & <$\alpha$> \\
$5.$    &     & $\rightarrow$ & <$\alpha$> <con> <$\alpha$> \\
$6.$     &    & $\rightarrow$ & <$\alpha$> <con> <$\alpha$> \\
    &    &  & <con> <$\alpha$> \\
$7.$ & <$\alpha$> & $\rightarrow$ & <pred> \\
$8.$    &     & $\rightarrow$ & <b\_pred> \\
$9.$    &     & $\rightarrow$ & <act\_pred> \\

$10.$ & <pred> & $\rightarrow$ & $x$ | $x$ $\circ$ $x$ | $x_1$ $\circ$ $x_2$\\
$11.$    &    & $\rightarrow$ & $\neg x$ | $\neg$($x$ $\circ$ $x$) | $\neg$ ($x_1$ $\circ$ $x_2$)\\

$12.$ & <b\_pred> & $\rightarrow$ & $x \geq u_1 \wedge x \leq u_2$\\
$13.$      &             & $\rightarrow$ & $ x_1 \geq x_2 \wedge x_1 \leq x_3$\\
$14.$       &             & $\rightarrow$ & $x < u_1 \vee x > u_2 $ \\
$15.$        &            & $\rightarrow$ & $ x_1 < x_2 \vee x_1 > x_3$\\

$16.$ & <act\_pred> & $\rightarrow$ & {\color{red}verb}(E) WITH E.<pred>$_1$ \\
&                    & & ... E.<pred>$_4$\\
$17.$ & <con> &  $\rightarrow$ & $\vee$ | $\wedge$ | X\\
$18.$ & $x$          &  $\rightarrow$ & var \\
$19.$     &       &  $\rightarrow$ & get\_var$_1$(var$_2$)\\
$20.$ & $var$ & $\rightarrow$ & $variable$ \\
$21.$ & & $\rightarrow$ & {\color{red}$noun$}
\end{tabular}
\caption{\small{The grammar used to generate the synthetic dataset.
Many constructs from general LTL have been reduced for use in our
problem domain.
Constructs created for our domain are highlighted in red.
The {\color{red}verb} predicate consists of all verbs in the
SpecNFS dataset whose operations cannot be defined by any Boolean
operation and thus require special functions. get\_VAR consists of
constructs that capture the "X of Y" statements in the SpecNFS and
NFS RFC domains.}}
\label{table:grammar}
\end{figure}

\section{Retriever Loss Function}
\label{appendix:sec:retriever_loss}
From \citet{flagembedding}, given $p$ and $q$ are the paired texts, $q`\in Q`$ is a negative example and $\tau$ is the temperature and $e$ represents encodings.

\begin{equation}
L = min. \sum_{(p,q)} -\log \frac{\exp^{\langle e_p,e_q\rangle/\tau} }{\exp^{\langle e_p,e_q\rangle/\tau} + \sum_{Q^{`}} \exp^{\langle e_p,e_{q^{`}}\rangle/\tau}}
\end{equation}

\section{Task Dataset Details}
\label{sec:appendix:dataset_details}

The details about each dataset are given below:

%----------------------------------------------------------------------
\textbf{NL2LTL:} The task was to parse specifications of NFS
operations into Linear Temporal Logic (LTL) statements.  Obtaining
training data is expensive, so we used the synthetic data generation
process outlined in Section~\ref{sec:dataset_creation}.  We describe
the grammar (Figure~\ref{table:grammar}) and generation process above,
in Section~\ref{appendix:sec:grammar}.  The OVCs in this task were
names of function predicates (verbs) and a subset of variables
(nouns), as indicated in red in our grammar.  We evaluated the model
(trained on synthetic data) on a test set of $100$ real NFS
requirement statements sampled from \citet{specnfs} and annotated with
their LTL expression by a group of experts (three university
professors and two graduate students).

% ----------------------------------------------------------------------
\textbf{NL2Code:} The task was to parse natural-language specifications to Python code.  Here, Python function names were the OVCs. We used the CoNaLa training and test cases from \citet{docprompting} for data augmentation.  To add expert knowledge, we used the documentation database provided with the dataset.  The expert lexicon was designed such that the keys were the first 200 characters from the documentation of the corresponding function, and the values were the function names.  We used the pre-defined test sets, as they contained functions unseen during training, thereby simulating OVCs.
% ----------------------------------------------------------------------

\textbf{NL2CMD:} The task was to parse natural-language queries into
Bash statements.  Command names such as \texttt{cat}, \texttt{ls},
etc., were used to simulate OVCs.  We used the TLDR dataset
from~\citet{docprompting} for our experiments.  As training data is
available, we augmented it with expert lexicons to train \name.  The
lexicon key was a concise natural-language statement describing the
command, created by taking the first line from the
tldr page for the command.  The lexicon's
value was the command name.  We tested with the predefined test sets
from the TLDR dataset, as it contains the OOD command names needed to
simulate OVC.

Table~\ref{table:examples} shows examples of natural-language input,
formal-language output, and the gold expert-knowledge lexicon from
each task dataset that was used in our
simulations. Table~\ref{tab:dataset_details} shows further details
about the number of data rows available/used for training and testing
\name.

\begin{table*}[!ht]
\small
\centering
\begin{tabular}{p{13cm}}
\Xhline{4\arrayrulewidth}
% %%%%% TASK 1 (NL2LTL)
\large{\texttt{\textbf{TASK 1}} - \texttt{NL2LTL}}
\\ \Xhline{4\arrayrulewidth}

\textbf{NL (NFS-RFC specification)} : If the current filehandle is not a regular file, an error will be returned to the client.
\\\Xhline{2\arrayrulewidth}
\textbf{Expert Knowledge} : $\langle$current filehandle $\rightarrow$  \texttt{cfh}$\rangle$, $\langle$A is a regular file $\rightarrow$ \texttt{is\_regular(A)}$\rangle$, $\langle$A returned $\rightarrow$ \texttt{return(A)}$\rangle$
\\\Xhline{2\arrayrulewidth}
\textbf{FL (Linear Temporal Logic)} : \texttt{ALWAYS(( !(is\_regular(cfh)) )$\implies$ ( return(error) ) )}\\
\Xhline{4\arrayrulewidth}

\large{\texttt{\textbf{TASK 2}} - \texttt{NL2CMD}}
\\ \Xhline{4\arrayrulewidth}
\textbf{NL (Instruction/Query)} : Delete a shared memory segment by id
\\
 \Xhline{1\arrayrulewidth}
\textbf{Expert Knowledge} : $\langle$Delete IPC (Inter-process Communication) resources.$\rightarrow$\texttt{ipcrm}$\rangle$\\
\Xhline{1\arrayrulewidth}
\textbf{FL (Bash statement)} : \texttt{ipcrm --shmem-id {{shmem\_id}}} \\
\Xhline{4\arrayrulewidth}
\large{\texttt{\textbf{TASK 3}} - \texttt{NL2Code}}
\\ \Xhline{4\arrayrulewidth}
\textbf{NL (Instruction/Query)} : divide the values with same keys of two dictionary `d1` and `d2`
\\ \Xhline{1\arrayrulewidth}
\textbf{Expert Knowledge} : $\langle$ class float([x])
Return a floating point number constructed from a number or string x. If the argument is a string, it should contain a decimal number, optionally preceded by a sign, and optionally e ... $\rightarrow$ \texttt{python.library.functions\#float} $\rangle$
\\ \Xhline{1\arrayrulewidth}
\textbf{Python} : \texttt{\{k: (float(d2[k]) / d1[k]) for k in d2\}} \\
\Xhline{4\arrayrulewidth}
\end{tabular}
\caption{\textbf{Evaluation Task Examples}: This table illustrates the
  NL input, target FL output, task example, and expert knowledge from
  the datasets for each task.}
\label{table:examples}
\end{table*}

\begin{table}[!ht]
    \centering
    \begin{tabular}{m{0.2\linewidth}|m{0.2\linewidth}|m{0.2\linewidth}|m{0.2\linewidth}}
    \Xhline{3\arrayrulewidth}
    \multirow{2}{*}{Dataset} & Retriever & Generator & \multirow{2}{*}{Test Set}\\
    & Train Set & Train Set & \\
    \Xhline{3\arrayrulewidth}
    % Synthetic & 120k & 120k & 2000\\
    % \hline
    NFS RFC & 120k & 120k & 100\\
    \hline
    TLDR & 8260 & 8260 & 928\\
    \hline
    Conala & 2135 & 2135 & 543\\
    \Xhline{3\arrayrulewidth}
    \end{tabular}
    \caption{Size of the different training and test sets for each the datasets used in evaluating the tasks.}
    \label{tab:dataset_details}
\end{table} 

\section{Reusability in Semantic Parsing}
\label{appendix:sec:reuse}

\begin{figure}[!ht]
\centering
\begin{subfigure}{.5\textwidth}
  \centering
  \includegraphics[width=\linewidth]{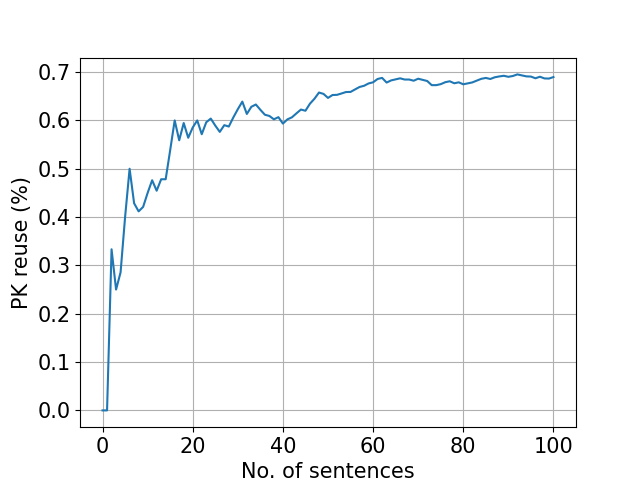}
  \caption{Reusability of\\ partial knowledge.}
  \label{fig:sub1}
\end{subfigure}%
\begin{subfigure}{.5\textwidth}
  \centering
  \includegraphics[width=\linewidth]{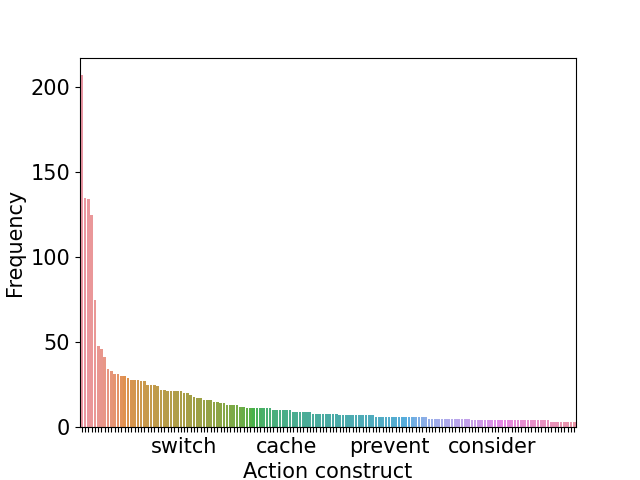}
  \caption{Distribution of actions in NFS-RFC}
  \label{fig:sub2}
\end{subfigure}
\caption{\small{\textbf{Observing reusability} : Analyses of the
    NFS specifications provides empirical support for the importance of
    modeling semantic parsing in an inference-time expert-knowledge
    setting.  The left half (a) shows the percentage
    of reuse that can happen when we collect the formal constructs in
    the expert knowledge base.  As we collect constructs from more
    sentences, the potential for reuse grows quickly.  The right half
    (b) shows the frequency of action-denoting words (possible OVCs) in the
    specification sentences.  There is a long tail of words that occur
    multiple times.  Knowledge learned about these once will likely be
    useful multiple times later.}}
%\sg{Font size of the axis labels \& ticks are too small to read. Also the caption needs further elaboration.}}
\label{fig:reuse}
\end{figure}

Figure \ref{fig:reuse} shows the potential for reusing expert
knowledge lexicons in an inference-time knowledge setting in the NFS
RFC domain. The diagram highlights the repetition of knowledge and the potential of reusability of knowledge about parsing OVCS.

% left part of the diagram shows the repetition of
% knowledge lexicons across the evaluation set; the right part shows the
% distribution of potential OVC reusability.

\section{Evaluation Setup and Metrics}
\label{appendix:sec:setup_and_eval}
Further details of the evaluation setup are given below:

\textbf{Retriever Training}: We utilize the small, base, and large variants of FlagEmbedding models~\citep{flagembedding}, which are sentence transformers~\citep{sbert} trained by BAAI. The dense retrievers are trained in an in-batch constrastive fashion using the loss function in Section \ref{appendix:sec:retriever_loss}. The retriever was trained for $5$ epochs using a batch size of $32$ and a learning rate of 1e--5.

\textbf{Generator Training}: We have two types of generators:
fine-tuning and few-shot. For fine-tuning, we train T5 and Code-T5
models. Each model was trained for $1$ to $10$ epochs with learning
rates of 1e--4 and batch size of $16$ to $64$. The models were trained
on an NVIDIA A6000 GPU for a total of approximately $24$ hours. For
few-shot learning, we used ChatGPT and GPT-4 in an incontext learning
setup by providing three incontext examples and generation temperature
of $0.01$.

Following are further details of the metrics used for evaluation:

\textbf{BLEU}: For measuring the quality of the overall parse by the
models, we \textsc{evaluate} the BLEU metric of the parse. We utilize the
evaluate library of Huggingface and calculate the
Sacrebleu~\citep{sacrebleu} metric for the NL2LTL and NL2CMD
datasets. For the NL2Code dataset, we use the BLEU metric system
and tokenization in the Docprompting paper. The formula for BLEU is as
follows:

\begin{equation}
    BLEU = bp.exp(\sum_{n=1}^{N}w_n\log p_n)
\end{equation}

where $bp$ is the brevity penalty, $w_n$ is the weight of each n-gram precision and $p_n$ is the precision of each n-gram and $N$ is the total number of n-gram precisions considered in calculating BLEU.

\textbf{OVC metrics}: To measure \name's ability to use expert
knowledge to parse the OVCs, we exclusively calculate metrics related
to the OVCs. This is done by first extracting the OVCs from the
generated text via pattern matching with the OVCs in the gold
lexicon. Then we calculate the unordered precision, recall and
F1-score using the following formulas.

\begin{equation}
    Precision = \frac{True\;Positive}{True\;Positive+False\;Positive}
\end{equation}

\begin{equation}
    Recall = \frac{True\;Positive}{True\;Positive+False\;Negative}
\end{equation}

\begin{equation}
    F1-Score = 2\times\frac{Precision\times Recall}{Precision+Recall}
\end{equation}

Note that all OVCs in our evaluation sets are by definition and construction unseen during training and the evaluation metrics above focus only on the OVC performance. Thus all gains observed on these metrics will primarily reflect on the models' abilities to handle expert knowledge effectively during inference time and not affect any knowledge memorized during training.  

\textbf{Gains}: \name is designed to be an improved form of feedback
system. In regular feedback systems for handling OVCs, an expert sits
at the output helm, reads through each parse, and then provides
corrective feedback for each erroneous OVC in the output. \name
reduces this amount of corrective feedback by reusing feedback from
the expert. In this way, manual effort is reduced. We formulate an
automatic way to measure this effort in the following manner: for each
parse, the expert needs to spend $1$ unit of work to read through the
whole parse. Then the expert needs to spend $1$ unit for each wrong
OVC in the output. This number can easily be calculated by measuring
the number of false negatives, i.e., OVCs missed in the output. The
sum of these two numbers measures the total amount of human effort
theoretically required to correct the OVCs. Hence the cost is:

\begin{equation}
    Cost = Reading\_Cost + Error\_Cost
\end{equation}

\section{Retriever Analysis}
\label{appendix:sec:retriever_ablations}

Table \ref{table:retriever_recall} shows the top-k recall on each
benchmark dataset for five different set of retrievers:
bm25~\citep{bm25}, off-the-shelf small and large  BGE sentence
transformers from FlagEmbeddings~\citep{flagembedding}, and the
fine-tuned versions of the aforementioned two models. We observe that bm25, being a
lexical retriever, has lowest performance on all tasks. Off-the-shelf
dense retrievers perform better, with the large variant doing better
than the smaller one except in the case of NL2LTL dataset. However,
the best performance is obtained with the fine-tuned retrievers. This
indicates the necessity of retriever training for our task.

\begin{table}[!ht]
\centering
\small
\begin{tabular}{m{0.1\linewidth}|m{0.2\linewidth}|m{0.07\linewidth}|m{0.07\linewidth}|m{0.07\linewidth}|m{0.07\linewidth}}
\Xhline{3\arrayrulewidth}
\multirow{2}{*}{\makecell[l]{Dataset}} & \multirow{2}{*}{\makecell[l]{Retriever}} & \multicolumn{4}{c}{Downstream}\\
\cline{3-6}
 & & R@1 & R@5 & R@10 & R@20 \\
\hline
 \multirow{7}{*}{\makecell[l]{NL2LTL}} & bm25 & 1.92 & 9.13 & 12.50 & 17.79\\
 \cline{2-6}
& Off-the-shelf & & & &\\
& -bge-small & 7.21 & 15.38 & 23.56 & 40.38\\
& -bge-large & 6.25 & 17.79 & 24.52 & 36.54\\
 \cline{2-6}
& Fine-tuned & & & &\\
& -bge-small & 11.54 & 30.77 & 43.75 & 47.12\\
& -bge-large & 14.42 & 36.06 & 44.71 & 50.00\\
\Xhline{2\arrayrulewidth}
\multirow{7}{*}{\makecell[l]{NL2CMD}} & bm25 & 1.72 & 5.28 & 11.53 & 24.25\\
 \cline{2-6}
& Off-the-shelf & & & &\\
& -bge-small & 40.19 & 57.97 & 63.25 & 68.64\\
& -bge-small(FT) & 46.34 & 63.69 & 68.75 & 
72.84\\
 \cline{2-6}
& Fine-tuned & & & &\\
& -bge-large & 41.27 & 59.27 & 65.30 & 69.94\\
& -bge-large(FT) & 52.59 & 68.86 & 72.20 & 75.32\\
\Xhline{2\arrayrulewidth}
 \multirow{7}{*}{\makecell[l]{NL2Code}} & bm25 & 0.34 & 1.35 &  2.59 & 5.29\\
\cline{2-6}
& Off-the-shelf & & & &\\
& -bge-small & 9.57 & 16.22 &  17.91 & 19.59\\
& -bge-small(FT) & 11.15 & 18.13 &  19.37 & 
21.51\\
\cline{2-6}
& Fine-tuned & & & &\\
& -bge-large & 9.23 & 16.67 &  18.81 & 20.27\\
& -bge-large(FT) & 11.37 & 18.69 & 20.38& 22.18\\

\Xhline{3\arrayrulewidth}
\end{tabular}
\caption{ Top-k recall performance (R@k) for different sets of retrievers on each benchmark dataset.}
\label{table:retriever_recall}
\end{table}

\section{Feedback Methods}
\label{appendix:sec:feedback}
Table \ref{tbl:generator_and_lexicon} compares \name using lexicons with two other methods: exemplar and \name using docs. 

\textbf{Examplar} refers to training the generator to use retrieved
pairs of input-output as context during retrieval augmented
generation. The idea comes from \citet{pasupat2021controllable} where
the authors retrieved similar input-output pairs for semantic
parsing. In our case, an off-the-shelf BGE large retriever is used to
retrieve (specification, code) pairs from the training set of NL2Code
to augment the training data with relevant context for
training. Afterwards, during testing, the test input is augmented with these
examples retrieved from the original training data.

The \textbf{\name with docs} idea comes from the paper
\citet{docprompting} where the models are conditioned to learn to use
documents to overcome the OOD problem. For the NL2Code task in Table
\ref{tbl:generator_and_lexicon}, we simply used the key of our lexicon
as the document. The retriever used is the fine-tuned BGE large used
in \name + lexicons use-case. The generator is conditioned only upon NL $\oplus$ key.

% Table \ref{table:retriever_recall} shows the top-k recall on each benchmark dataset for five different set of retrievers: bm25\cite{bm25}, off-the-shelf small and large  BGE sentence transformers from FlagEmbeddings, and the fine-tuned versions of these two models. We observe that bm25, being a lexical retriever, has lowest performance on all tasks. Off-the-shelf dense retrievers perform better, with the large variant doing better than the smaller one except in the case of NL2LTL dataset. However, the best performance is obtained with the fine-tuned retrievers. This indicates the necessity of retriever training for our task.

\section{Gains Analysis}
\label{appendix:sec:gains}

Table \ref{table:gains} shows the gains calculated as per our metric
defined in Section \ref{sec:benchmarking_setup} and and elaborated in
Section \ref{appendix:sec:setup_and_eval}. For each dataset in the
fine-tuned setting, we select the model with the highest difference in
performance on OVCs. Then we calculate the reading cost and feedback
cost. Reading cost is equal to the size of the dataset while feedback
cost is equivalent to the false negative among the OVCs. This is then
added to find the total. The percentage reduction between baseline and
\name is then calculated.We see $24.3$\% savings for NL2LTL, $5.6\%$
for NL2Code, $11.9\% $for NL2CMD, and $32.9\%$ for few-shot learning
in NL2LTL. This shows the potential of \name in reducing human
effort.

\begin{table}[!ht]
    \centering
    \begin{tabular}{m{0.11\linewidth}|m{0.15\linewidth}|m{0.1\linewidth}|m{0.1\linewidth}|m{0.1\linewidth}|m{0.1\linewidth}}
    \Xhline{3\arrayrulewidth}
    \textbf{Dataset} & Model & Reading & Error & Total & Reduction \\
    \Xhline{2\arrayrulewidth}
    \hline
    \multirow{2}{*}{NL2LTL} & CT5-B & 100 & 180 & 280 & \\
    & +\name & 100 & 112 & 212 & -24.3\%\\
    \Xhline{2\arrayrulewidth}
    \hline
    \multirow{2}{*}{NL2Code} & T5-B & 543 & 345 & 888 & \\
    & +\name & 543 & 295 & 838 & -5.6\%\\
    \Xhline{2\arrayrulewidth}
    \hline
    % \multirow{2}{*}{NL2CMD} & CT5-S & 928 &  631 & 1559 & \\
    % & +\name & 100 &  59 & 159 & -11.9\%\\
    \multirow{2}{*}{NL2CMD} & CT5-S & 928 &  878 & 1806 & \\
    & +\name & 928 &  583 & 1511 & -16.3\%\\
    \Xhline{2\arrayrulewidth}
    \hline
    \multirow{2}{*}{Few-shot} & GPT-4 & 100 &  137 & 237 & \\
    & +\name & 100 &  59 & 159 & -32.9\%\\
    \Xhline{3\arrayrulewidth}
    \end{tabular}
    \caption{Cost analysis of \name compared to a regular feedback system using our defined ``gain'' metric. Each model is selected based on the maximum possible gain. This gain is shown as reduction in this table for each corresponding \name.}
    \label{table:gains}
\end{table}

\section{Error Analysis}
\label{appendix:sec:error_analysis}
Table \ref{table:error_analysis} shows the detailed error analysis on
the OVCs for the NL2LTL and NL2Code datasets. We selected the best
models for each dataset, T5-Base and Code-T5 Base respectively, and analyzed the $25$ most
erroneous outputs. Four categories were identified as common errors:
expert lexicon is available but not used, incorrect expert lexicon is
used, expert lexicon is available in the partial knowledge for
retrieval but not retrieved, and OVC was hallucinated even though
there were no corresponding expert lexicon in the input.

From the NL2LTL portion of the analysis, we observe that most errors
corresponded to ``hallucinating expert knowledge''. This shows that
the model tries to generalize without an expert lexicon, which is
against our intended use case. The model also has a high probability
of not using an expert lexicon or using an incorrect one.

For the NL2Code analysis, we omitted the ``hallucinated EK'' due to
the large number of possible functions that the model can generate
that may be an OVC. Also, due to the sparsity of OVCs in the dataset,
we only select instances where an OVC is present. We observe that a
large portion of the errors are when the model is not using an expert
lexicon. This hints at the difficulty of using lexicons to resolve
OVCs on this benchmark.

\begin{table}[!ht]
\centering
\small
\begin{tabular}{m{0.25\linewidth}|m{0.1\linewidth}|m{0.1\linewidth}}
\Xhline{3\arrayrulewidth}
\multirow{2}{*}{\textbf{Error Class}} & \multicolumn{2}{c}{Percentage}\\
\cline{2-3}
& NL2LTL & NL2Code\\
\hline
EK not used & 23.3 & 56.0\\
\hline
Incorrect EK used & 23.3 & 16.0\\
\hline
EK not retrieved & 6.7 & 28.0\\
\hline
Hallucinated EK & 46.7 & -\\
\Xhline{3\arrayrulewidth}
\end{tabular}
\caption{Error analysis of \name.}
\label{table:error_analysis}
\end{table}

\end{document}